\documentstyle[preprint,pre,aps]{revtex} 
\begin{document}                                     
\draft
\title{Modeling Multi-Lane Traffic Flow with Queuing Effects}
\author{Dirk Helbing}
\address{II. Institute of Theoretical Physics, University of
Stuttgart, 70550 Stuttgart, Germany}
\maketitle
\vfill                                                    
\begin{abstract}
On the basis of assumptions about the behavior of driver-vehicle units
concerning acceleration, deceleration, overtaking, and lane-changing
maneuvers, a gas-kinetic traffic model for uni-directional multi-lane
freeways is constructed. Queuing effects are explicitly taken into
account in an overall manner. The resulting model is a generalization of
Paveri-Fontana's Boltzmann-like traffic model and allows the derivation
of macroscopic traffic equations for interacting lanes, including
velocity equations. The related effective macroscopic traffic model for
the total freeway cross-section is also derived. It
provides corrections with respect to
previous traffic models, but agrees with them in special cases.\\[5mm]
Key Words: Multi-lane traffic; Queuing; Gas-kinetic traffic model;
Macroscopic traffic equations; Paveri-Fontana model.
\end{abstract}                                                
\pacs{PACS numbers: 51.10.+y, 89.40.+k, 47.90.+a, 34.90.+q}
\vfill
\clearpage

\section{Introduction}
Apart from microscopic traffic models, 
in the last decades a number of interrelated macroscopic traffic models have
been proposed 
\cite{LiWhi55,Ri56,Pay71,Pay79b,Kue84,KueRoe91,KeKo93,KeKo94,Hel95a,Hel95b,Hel96}.
The motivations for developing these were
\begin{itemize}
\item to describe and understand the instabilities of traffic flow
\cite{Kue84,KueRoe91,KeKo93,KeKo94,Hel96,KueBe93,Hel95c}, 
\item to optimize traffic flow by means of on-line speed-control
systems \cite{Kue87,KueKr92,Hel95d},
\item to make short-term forecasts of traffic volumes for re-routing
measures \cite{HiReWe93,Hil95,Hil96},
\item to calculate the average travel times, fuel consumption, and
vehicle emissions in dependence of traffic volume 
\cite{Pay79b,Hel97},
\item to predict the effects of additional roads or lanes 
\cite{Hel97,Bra68,Bas92,Kno93}. 
\end{itemize} 
Most of these models are restricted to uni-directional freeway traffic and 
treat the different lanes of a road in an overall manner, i.e. 
like one lane with higher capacity and possibilities for overtaking.
However, this kind of simplification is clearly not applicable 
if there is a disequilibrium between neighboring lanes. 
Therefore, some researchers carried out empirical
investigations of the observed density oscillations between neighboring lanes
or proposed models for their mutual influences 
\cite{GaHeWe62,MuPi71,MuHs71,Ror76,MaNa81,MiBe84}.
\par
However, these are phenomenological models which
treat inter-lane interactions in a rather heuristic
way. Moreover, most of them base on the simple traffic flow model of 
Lighthill and Whitham which assumes average velocity on each lane
to be in equilibrium with density. This assumption is not very well justified,
especially for unstable traffic which is characterized by evolving
``phantom traffic jams'' or stop-and-go waves
\cite{Kue84,KueRoe91,KeKo93,KeKo94,Hel96,KueBe93,Hel95c}.
It is also questionable for lane mergings or on-ramp traffic where
frequently a disequilibrium occurs.
However, instabilities or disequilibria may decrease the freeway capacity
considerably. 
\par
An alternative approach including a phenomenological velocity equation
has been proposed by Michalopoulos et al.
\cite{MiBe84}. It bases on Payne's model \cite{Pay71,Pay79b} which 
has been severely criticized for several reasons
\cite{Kue84,Hel96,HaHu79,Pay79,Pap83,Cre85,Smu87,RaLi87}.
Therefore, we will derive a consistent macroscopic multi-lane model 
from a {\em gas-kinetic} level of description.
This is related to Paveri-Fontana's
Boltzmann-like approach (cf. Sec. II), but explicitly takes
into account overtaking and lane-changing maneuvers. An extension
to a gas-kinetic traffic model for
several driving styles (e.g. aggressive,
timid) or vehicle types (e.g. cars, trucks) is easily 
possible (cf. Sec.~\ref{Disti}).
\par
On the way of constructing gas-kinetic traffic equations we face the problem
of vehicle queues (platoons) forming at medium and high traffic volumes
\cite{Dag75}. Several solutions have been suggested to cope with this.
Most of them base on gas-kinetic
ideas, but non of them is fully satisfactory, yet. The approach of Andrews 
\cite{And1,And2,And3,And4}
mainly provides results for stationary traffic situations.
Beylich's model \cite{Be79,Be81} is so complicated that the corresponding 
{\em macroscopic} traffic equations are not any more derivable or at least 
not suited for efficient traffic simulations. For this reason, this paper 
does not distinguish queues of different lengths (like Beylich does) but only
freely moving and impeded (queued) vehicles. 
This approach has some relation to the one of Lampis
\cite{Lam78}, but it removes its inconsistencies.
\par
The resulting Boltzmann-like model allows a systematic derivation of
macroscopic traffic equations not only for the vehicle densities 
on the different lanes, but also 
for the associated average velocities (cf.
Sec. III). In Sec. IV it is demonstrated how effective macroscopic traffic
equations for the {\em total} cross-section of a uni-directional road can be 
obtained from the equations for single lanes. 
Due to different legal regulations, the traffic dynamics on American
freeways is different from that on European ones (which will be called
{\em ``autobahns''} in accordance with K\"uhne et al. \cite{Kue87,KueRoe91})
(cf. Sec. II).
\par
A summary and outlook is presented in Sec. V.

\section{Boltzmann-like multi-lane theory}

The first Boltzmann-like (gas-kinetic) model was proposed by Prigogine and
co-workers \cite{PrAn60,Pri61,PrHe71}. However, Paveri-Fontana \cite{Pa75} 
has pointed out that this model has some peculiar 
properties. Therefore, Paveri-Fontana proposed an improved model that
overcomes most of the short-comings of 
Prigogine's approach. Nevertheless, his model still treats
the lanes of a multi-lane road in an overall manner and
does not take into account queuing effects. For this reason, an extended
Paveri-Fontana-like model will now be constructed.
\par
Let us assume that the motion of an individual vehicle $\alpha$ can be
described by several variables like its {\em lane} $i_\alpha(t)$,
its {\em place} $r_\alpha(t)$, its {\em actual velocity}
$v_\alpha(t)$, and its {\em desired velocity} $v_{0\alpha}(t)$
in dependence of time $t$.
The {\em phase-space density}
$\hat{\rho}_i(r,v,v_0,t)$
is then determined by the mean number $\Delta 
n_i(r,v,v_0,t)$ of vehicles on {\em lane} $i$ that are at a
place between $r - \Delta r/2$ and $r + \Delta r/2$, 
driving with a velocity between $v - \Delta v/2$ and $v + \Delta v/2$, and
having a desired velocity between $v_0 - \Delta v_0/2$ and $v_0 + \Delta v_0/2$
at time $t$:
\begin{eqnarray}
 \hat{\rho}_i(r,v,v_0,t) 
 &=& \frac{\Delta n_i(r,v,v_0,t)}{\Delta r \, \Delta v \, \Delta v_0} 
 \nonumber \\
 & & \hspace*{-1.5cm} =
  \frac{1}{\Delta r \, \Delta v \, \Delta v_0} \sum_\alpha 
 \delta_{ii_\alpha(t)} 
 \!\!\!\!\!\int\limits_{r-\Delta r/2}^{r+\Delta r/2}\!\!\!\!\! dr'
 \delta( r' - r_\alpha(t))
 \!\!\!\!\!\int\limits_{v-\Delta v/2}^{v+\Delta v/2}\!\!\!\!\! dv'
 \delta( v' - v_\alpha(t)) 
 \!\!\!\!\!\int\limits_{v_0-\Delta v_0/2}^{v_0+\Delta v_0/2}\!\!\!\!\! dv'_0 \,
  \delta( v'_0 - v_\alpha^0(\!\not t)) \, . 
 \nonumber \\
 & & 
\end{eqnarray}
Here, $\Delta r$, $\Delta v$, and $\Delta v_0$ are small
intervals. $\delta_{ij}$ denotes the Kronecker symbol and
$\delta(x-y)$ Dirac's delta function. The notation ``$\not t$'' 
indicates that a time-dependence only occurs in exceptional cases.
Lane numbers $i$ are counted in increasing order from the
right-most to the left-most lane, but in Great Britain and Australia
the other way round. (For Great Britain
and Australia ``left'' and ``right'' must always be interchanged.) 
\par
The phase-space density $\hat{\rho}_i(r,v,v_0,t)$ can be splitted
into a term $\hat{\rho}_{i1}(r,v,v_0,t)$ describing vehicles that can
move freely and a term $\hat{\rho}_{i2}(r,v,v_0,t)$ delineating impeded vehicles
that have to move slower than desired since they are queuing behind other
vehicles:
\begin{equation}
 \hat{\rho}_i(r,v,v_0,t) = \hat{\rho}_{i1}(r,v,v_0,t) +\hat{\rho}_{i2}(r,v,v_0,t)
 \, .
\end{equation}
Introducing the {\em proportion $c_i(r,v,v_0,t)$ of freely moving vehicles}
by
\begin{equation}
 c_i(r,v,v_0,t)\hat{\rho}_i(r,v,v_0,t) = \hat{\rho}_{i1}(r,v,v_0,t) \, ,
\end{equation}
we have the relations
\begin{equation}
 0 \le c_i(r,v,v_0,t) \le 1
\end{equation}
and
\begin{equation}
 \hat{\rho}_{i2}(r,v,v_0,t) = [ 1 - c_i(r,v,v_0,t) ]
 \hat{\rho}_i(r,v,v_0,t) \, .
\end{equation}
\par
Now, we utilize the fact that, due to the conservation of the number of
vehicles, the phase-space density $\hat{\rho}_i(r,v,v_0,t)$ on lane $i$
obeys the {\em continuity equation} \cite{Hel97,Pa75,AlBe78}
\begin{eqnarray}
 \frac{\partial \hat{\rho}_i}{\partial t} + 
 \frac{\partial}{\partial r} ( \hat{\rho}_i v )
 + \frac{\partial}{\partial v} ( \hat{\rho}_i f_i^0 )
 &=& \left( \frac{\partial \hat{\rho}_i}{\partial t} \right)_{\rm ad}
 \!\! + \left( \frac{\partial \hat{\rho}_i}{\partial t} \right)_{\rm vd}
 \!\! + \left( \frac{\partial \hat{\rho}_i}{\partial t} \right)_{\rm int}
 \!\! + \left( \frac{\partial \hat{\rho}_i}{\partial t} \right)_{\rm lc}
 \nonumber \\[2mm]
 &+& \hat{\nu}_i^+(r,v,v_0,t) - \hat{\nu}_i^-(r,v,v_0,t) \, .
\label{Boltz1}
\end{eqnarray}
The second and third term describe temporal changes of the phase-space
density $\hat{\rho}_i(r,v,v_0,t)$ due to changes $dr/dt = v$
of place $r$ and due to acceleration $f_i^0$, respectively.
We will assume that the proportion $c_i(r,v,v_0,t)$ of freely moving
vehicles accelerates to their desired velocity $v_0$ with a
certain {\em relaxation time} $T$ so that we have the {\em acceleration law}
\begin{equation}
 f_i^0(r,v,v_0,t) = \frac{c_i}{T} (v_0 - v) \, .
\label{acc}
\end{equation}
The terms on the right-hand side of equation (\ref{Boltz1}) reflect changes
of phase-space density $\hat{\rho}_i(r,v,v_0,t)$ due to 
{\em discontinuous} changes of desired velocity $v_0$, actual velocity $v$, 
or lane $i$. $\nu_i^+(r,v,v_0,t)$ and $\nu_i^-(r,v,v_0,t)$ are the {\em rates
of vehicles entering and leaving the road} at place $r$. They are only
different from zero for merging lanes at entrances and exits respectively.
\par
The term
\begin{equation}
 \left( \frac{\partial \hat{\rho}_i}{\partial t} \right)_{\rm ad} 
 = \frac{\tilde{\rho}_i(r,v,t)}{T_{\rm r}} [ \hat{P}_{0i}(v_0;r,\!\not t)
 - P_{0i}(v_0;r,t) ] \, ,
\label{adj}
\end{equation}
where 
\begin{equation}
 \tilde{\rho}_i(r,v,t) = \int dv_0 \, \hat{\rho}_i(r,v,v_0,t)
\end{equation}
is  a {\em reduced phase-space density} and $T_{\rm r} \approx 1$\,s is
about the {\em reaction time}, describes an adaptation of the {\em actual
distribution of desired velocities} $P_{0i}(v_0;r,t)$ 
to the {\em reasonable distribution of desired velocities}
$\hat{P}_{0i}(v_0;r,\!\not t)$ without any related change of actual velocity 
$v$. 
\par
For the reasonable distribution of desired velocities we will assume the
functional dependence
\begin{equation}
 \hat{P}_{0i}(v_0;r,\!\not t) = {\textstyle\frac{1}{\sqrt{2\pi
 \hat{\theta}_{0i}}}} \mbox{e}^{-[v_0 - \hat{V}_{0i}]^2/[2 \hat{\theta}_{0i}]}
\end{equation}
which corresponds to a normal distribution and is empirically
well justified \cite{Pam55,MuPi71,Hel97}. The mean value 
$\hat{V}_{0i} = \hat{V}_{0i}(r,\!\not \!t)$ and variance $\hat{\theta}_{0i}
= \hat{\theta}_{0i}(r,\!\not \!t)$ of $\hat{P}_{0i}(v_0;r,\!\not \!t)$ depend
on road conditions and speed limits. 
Since European autobahns usually do not have speed limits (at least
in Germany), $\hat{\theta}_{0i}$ is larger for these than for American 
freeways. In addition, on European autobahns
$\hat{V}_{0i}$ increases with increasing lane number $i$ since
overtaking is only allowed on the left-hand lane. 
\par
The term
\begin{equation}
 \left( \frac{\partial \hat{\rho}_i}{\partial t} \right)_{\rm vd}
 = \frac{\partial^2}{\partial v^2} \left( \hat{\rho}_i
 \frac{A_i v^2}{T} \right)  
\label{VD}
\end{equation} 
with a dimensionless, density-dependent function $A_i$ describes
a kind of {\em ``velocity diffusion''} which takes into account
individual fluctuations of velocity due to imperfect driving.
For a detailled discussion of this term cf. Refs.~\cite{Hel96,Hel97}. 
It provides
contributions to the dynamical variance equations and equations
for higher moments, but neither to the density nor to the
velocity equations \cite{Hel97}.
\par
Before we specify the Boltzmann-like interaction term $(\partial \hat{\rho}_i
/ \partial t)_{\rm int}$ and the lane-changing term $(\partial \hat{\rho}_i
/\partial t)_{\rm lc}$ we will discuss some preliminaries. For reasons of
simplicity we will only treat vehicle interactions within the {\em same} lane 
as {\em direct pair interactions,} i.e. in a Boltzmann-like manner
\cite{QSoz}. Lane-changing maneuvers of impeded vehicles that want to escape a queue
(i.e. leave and overtake it) may depend on interactions of up to six vehicles
(the envisaged vehicle, the vehicle directly in front of it, and up to
two vehicles on both neighboring lanes which may prevent overtaking
if they are too close). Therefore, we will treat lane-changing maneuvers
in an overall manner by specifying overtaking probabilities and waiting times of 
lane-changing maneuvers (which corresponds to a {\em mean-field approach},
cf. Ref.~\cite{QSoz}).
These probabilities and waiting times dependent on the
vehicle densities and maybe also on other quantities.

\subsection{Overtaking}

Let, for example, $P_i^+$ denote the probability that a slower vehicle
could be immediately overtaken on the left-hand lane and $P_i^-$ the analogous
probability for the right-hand lane. 
In addition, let $P_{i_0} \equiv P_{i_0}(r,\!\not t)$ be the proportion
of driver-vehicle units that desire to move on lane $i_0$ (which corresponds
to the proportion of vehicles that {\em actually} 
moves on lane $i_0$ at small vehicle densities). Then, 
\begin{equation}
 p_i^> \equiv p_i^>(r,\!\not t) = \sum_{i_0(> i)} P_{i_0}(r, \!\not t)
\end{equation}
is the proportion of vehicles on lane $i$ which would prefer to overtake
on or change to the left-hand lane $i+1$, whereas 
\begin{equation}
 p_i^< \equiv p_i^<(r,\!\not t) = \sum_{i_0(< i)} P_{i_0}(r, \!\not t)
\end{equation}
is the proportion of vehicles which would prefer to overtake on or change to
the right-hand lane $i-1$. For American freeways we 
will now calculate the probability $p_i^+$ [$p_i^-$]
with which vehicles on lane $i$
are immediately overtaken on lane $i+1$ [lane $i-1$] by vehicles that
also moved on lane $i$ beforehand. Since $P_i^+(1-P_i^-)$
[$P_i^-(1 - P_i^+)$] is the probability that only lane $i+1$
[lane $i-1$] allows overtaking, and
$P_i^+ P_i^-$ is the probability that {\em both} neighboring lanes are free, we
obtain
\begin{eqnarray}
 p_i^+ &=& c_i \{ P_i^+(1 - P_i^-) + [p_i^> + {\textstyle\frac{1}{2}} 
 ( 1 - p_i^> - p_i^<)] P_i^+ P_i^- \} \nonumber \\
 &=&  c_i [P_i^+(1 - P_i^-)
 + {\textstyle\frac{1}{2}} ( 1 + p_i^> - p_i^<) P_i^+ P_i^- ]
\label{pi1}
\end{eqnarray}
and
\begin{eqnarray}
 p_i^- &=& c_i \{ P_i^-(1 - P_i^+) + [p_i^< + {\textstyle\frac{1}{2}} 
 ( 1 - p_i^> - p_i^<)] P_i^+ P_i^- \} \nonumber \\
 &=& c_i [ P_i^-(1 - P_i^+)
 + {\textstyle\frac{1}{2}} ( 1 + p_i^< - p_i^>) P_i^+ P_i^- ]
\label{pi2}
\end{eqnarray}
for inner lanes. Here, we have taken into account that only the proportion
$c_i$ of freely moving vehicles is able to {\em immediately} overtake.
Moreover, we have assumed that driver-vehicle units 
change towards their 
desired lane $i_0$ if both neighboring lanes allow overtaking. If already
being on the desired lane $i_0 = i$ (which is the case with probability
$1 - p_i^> - p_i^<$), they choose each neighboring lane with probability
$1/2$. If only one lane is free, this chance is taken irrespective of the
desired lane $i_0$. From (\ref{pi1}) and (\ref{pi2}) follows that the 
total probability of immediate overtaking is
\begin{equation}
 p_i^+ + p_i^- = c_i [ 1 - (1 - P_i^+)(1 - P_i^-)] \, .
\end{equation}
\par
Since, on outer lanes, only one lane is available for overtaking maneuvers, we
have
\begin{equation}
 p_i^+ = 0 \, , \qquad p_i^- = c_i P_i^-
\end{equation}
on the left-most lane and
\begin{equation}
 p_i^- = 0 \, , \qquad p_i^+ = c_i P_i^+ 
\end{equation}
on the right-most lane.
\par
Of course, other specifications of $p_i^+$ and $p_i^-$ are also possible.
In Europe, for example, overtaking on the right-hand lane is prohibited
for free traffic flow (with less than 30 vehicles per kilometer and lane)
so that
\begin{equation}
 p_i^+ = c_i P_i^+ \, , \qquad p_i^- = 0
\end{equation}
on inner lanes and the right-most lane, but
\begin{equation}
 p_i^+ = 0 \, , \qquad p_i^- = 0
\end{equation}
on the left-most lane. However, for congested traffic (with an average velocity
less than 80 kilometers per hour) vehicles are also allowed to pass slower
vehicles on the right-hand lane so that the situation is similar
to American highways, then.
\par
We are now ready to specify the Boltzmann-like interaction term. For
not too large vehicle densities it can be written in the form \cite{Hel97}
\begin{mathletters}\label{Boltz2}\begin{eqnarray}
 \left( \frac{\partial \hat{\rho}_i}{\partial t} \right)_{\rm int}
 &=& \sum_{i'} \int dv' \!\!\int\limits_{w < v'} \!\! dw \int dw_0 \,
 W_2(v,i|v',i';w,i') \hat{\rho}_{i'}(r,v',v_0,t) \hat{\rho}_{i'}(r,w,w_0,t)
 \label{Boltz2a} \\
 &-& \sum_{i'} \int dv' \!\!\int\limits_{w < v} \!\! dw \int dw_0 \,
 W_2(v',i'|v,i;w,i) \hat{\rho}_i(r,v,v_0,t) \hat{\rho}_i(r,w,w_0,t) \, .
\label{Boltz2b}
\end{eqnarray}\end{mathletters}
Term (\ref{Boltz2a}) describes an increase of phase-space density 
$\hat{\rho}_i(r,v,v_0,t)$ by interactions of a vehicle with actual velocity
$v'$ and desired velocity $v_0$ on line $i'$ 
with a slower vehicle with actual velocity
$w < v'$ and desired velocity $w_0$ causing the former vehicle to 
change its velocity to $v \ne v'$ or its lane to $i\ne i'$. The frequency
of such interactions is proportional to the phase-space density 
$\hat{\rho}_{i'}(r,w,w_0,t)$ of hindering vehicles and the phase-space density
$\hat{\rho}_{i'}(r,v',v_0,t)$ of vehicles which can be affected by slower
vehicles. Analogously, term (\ref{Boltz2b}) describes 
a decrease of phase-space density $\hat{\rho}_i(r,v,v_0,t)$ 
by interactions of a vehicle with actual velocity
$v$ and desired velocity $v_0$ on line $i$ 
with a slower vehicle with actual velocity
$w < v$ and desired velocity $w_0$ causing the former vehicle to 
change its velocity to $v' \ne v$ or its lane to $i'\ne i$. Since the
interaction is assumed not to influence the desired velocities $v_0$,
$w_0$, the interaction rate $W_2$ is independent of these. However, the
interaction rate $W_2(v',i'|v,i;w,i)$ is proportional to the relative
velocity $|v-w|$ of approaching vehicles. Therefore, we have the following
relation:
\begin{mathletters}\label{intrate}\begin{eqnarray}
 W_2(v',i'|v,i;w,i) &=& p_i^+ |v-w|
 \delta_{i'(i+1)} \delta(v' - v) \label{inta} \\
 &+& p_i^- |v-w| \delta_{i'(i-1)}
 \delta(v' - v) \label{intb} \\
 &+& (1 - p_i) |v - w|
 \delta_{i'i} \delta(v' - w) \, . \label{intc} 
\end{eqnarray}\end{mathletters}
Term (\ref{inta}) describes an {\em undelayed overtaking} on 
lane $i' = i+1$ without any change of velocity ($v' = v$)
by vehicles which would be hindered by slower 
vehicles on lane $i$. Analogously, term (\ref{intb}) reflects undelayed
overtaking maneuvers on lane $i' = i-1$. Term (\ref{intc})
delineates situations where a vehicle cannot be immediately
overtaken by a faster vehicle so that the latter must stay on
the same lane ($i' = i$) and decelerate to the velocity $v' = w$ of the
hindering vehicle. 

\subsection{Lane-Changing}

We come now to the specification of the lane-changing term $(\partial
\hat{\rho}_i/\partial t)_{\rm lc}$. This has the form of a master equation:
\begin{mathletters}\label{spont}\begin{eqnarray}
 \left( \frac{\partial \hat{\rho}_i}{\partial t}\right)_{\rm lc}
 &=& \sum_{i'(\ne i)} W_1(i|i') \hat{\rho}_{i'}(r,v,v_0,t) \label{spona} \\
 &-& \sum_{i'(\ne i)} W_1(i'|i) \hat{\rho}_{i}(r,v,v_0,t) \, . \label{sponb}
\end{eqnarray}\end{mathletters}
Term (\ref{spona}) describes an increase of phase-space density
$\hat{\rho}_{i}(r,v,v_0,t)$ due to changes from lane $i' \ne i$ to lane $i$
by vehicles with actual velocity $v$ and desired velocity $v_0$. 
The frequency of lane-changing
maneuvers is proportional to the phase-space density
$\hat{\rho}_{i'}(r,v,v_0,t)$ of vehicles which may be interested in
lane-changing. Analogously, term (\ref{sponb}) reflects changes from lane $i$
to another lane $i'$ causing a decrease of $\hat{\rho}_{i}(r,v,v_0,t)$. 
For the corresponding rate $W_1(i'|i)$ of lane-changing maneuvers we assume
\begin{mathletters}\label{w}\begin{eqnarray}
 W_1(i'|i) &=& \frac{1}{\tilde{T}_i^+} [ 1 - c_{i}(r,v,v_0,t) ]
 \delta_{i'(i+1)} \label{wa} \\
 &+& \frac{1}{\tilde{T}_i^-} [ 1 - c_{i}(r,v,v_0,t) ]
 \delta_{i'(i-1)} \label{wb} \\
 &+& \frac{p_i^>}{\hat{T}_i^+} r_i \delta_{i'(i+1)} \label{wc} \\
 &+& \frac{p_i^<}{\hat{T}_i^-} r_i \delta_{i'(i-1)} \, . \label{wd} 
\end{eqnarray}\end{mathletters}
Term (\ref{wa}) describes lane changes by vehicles which are able to
escape a queue on the left-hand lane $i' = i+1$. $\tilde{T}_i^+$ is the
{\em average waiting time} until this is possible. Analogously, term
(\ref{wb}) reflects {\em delayed overtaking maneuvers} by impeded vehicles
on the right-hand lane $i' = i-1$. For free traffic flow on
European autobahns we must set
\begin{equation}
 \frac{1}{\tilde{T}_i^-} = 0
\end{equation}
since overtaking is only allowed on the left-hand lane $i' = i+1$.
\par
Term (\ref{wc}) delineates {\em spontaneous lane changes} to 
lane $i' = i+1$ by vehicles that desire to move on lane $i_0 > i$.
Analogously, (\ref{wd}) corresponds to changes to lane $i' = i-1$
by vehicles that desire
to move on lane $i_0 < i$. $\hat{T}_i^+$ and $\hat{T}_i^-$ denote the
related average waiting times. 
\par
For American freeways we have 
\begin{equation}
 r_i \equiv r_i(r,v,v_0,t) = c_i(r,v,v_0,t) \, ,
\end{equation}
since lane changing by queuing vehicles corresponds to delayed overtaking
which is described by terms (\ref{wa}) and (\ref{wb}). For 
free traffic flow on European autobahns
we have $p_i^> = 0$ due to a regulation prescribing to use
the right-most lane if possible. This also applies to queuing
vehicles so that, additionally,
\begin{equation}
 r_i = 1 \, .
\end{equation}
\par
We close with some remarks regarding the order of magnitude of
the waiting times $\tilde{T}_i^\pm$ and $\hat{T}_i^\pm$. First of all, they
depend on the vehicle density on the neighboring lane $i\pm 1$.
Moreover, for the left-most lane we must set 
\begin{equation}
 \frac{1}{\tilde{T}_i^+} = 0 \qquad \mbox{and} \qquad 
 \frac{1}{\hat{T}_i^+} = 0 \, ,
\end{equation}
whereas for the right-most lane we always have 
\begin{equation}
 \frac{1}{\tilde{T}_i^-} = 0 \qquad \mbox{and} \qquad 
 \frac{1}{\hat{T}_i^-} = 0 \, . 
\end{equation}
On American freeways, the waiting times $\hat{T}_i^\pm$
for spontaneous lane changing are greater than the waiting times
$\tilde{T}_{i\pm 1}^\mp$ for delayed overtaking since overtaking is usually carried out
as soon as possible, whereas changing towards the desired lane is done at one's
convenience (if one does not intend to leave the next exit), i.e.
\begin{equation}
 \frac{1}{\hat{T}_i^\pm} < \frac{1}{\tilde{T}_{i\pm 1}^\mp} \, .
\end{equation}
On European autobahns, the waiting time $\hat{T}_i^-$ for spontaneous
lane changing (to the right-hand lane) 
is even {\em much} larger than the waiting time
$\tilde{T}_{i-1}^+$ for delayed overtaking (on the left-hand lane) 
since the right-hand
lane is only accepted when it offers really large gaps. (Nobody wants to 
move on the slower lane.)
\par
Defining the abbreviations
\begin{equation}
 \frac{1}{T_i^+} = \frac{1 - c_i}{\tilde{T}_i^+} + \frac{p_i^> r_i}
 {\hat{T}_i^+} 
\end{equation}
and
\begin{equation}
 \frac{1}{T_i^-} = \frac{1 - c_i}{\tilde{T}_i^-} + \frac{p_i^< r_i}
 {\hat{T}_i^-} \, , 
\end{equation}
we can rewrite expression (\ref{w}) in the simple form
\begin{equation}
 W_1(i'|i) = \frac{1}{T_i^+} \delta_{i'(i+1)} + \frac{1}{T_i^-}
 \delta_{i'(i-1)} \, .
\label{w1}
\end{equation}

\subsection{Distinction of different vehicle types}\label{Disti}

The gas-kinetic traffic model developed above can be easily generalized to
cases where effects of different vehicle types (e.g. cars, trucks)
or driving styles (e.g. aggressive, timid) are to be investigated. Then,
the phase-space density $\hat{\rho}_i(r,v,v_0,t)$ can be splitted into
partial phase-space densities $\hat{\rho}_i^a(r,v,v_0,t)$ describing
driver-vehicle units of {\em type} $a$:
\begin{equation}
 \hat{\rho}_i(r,v,v_0,t) = \sum_a \hat{\rho}_i^a(r,v,v_0,t) \, .
\end{equation}
The corresponding gas-kinetic equations have an analogous form to
(\ref{Boltz1}) with (\ref{acc}), (\ref{adj}), (\ref{VD}), (\ref{Boltz2}),
and (\ref{spont}). 
They read
\begin{eqnarray}
 \frac{\partial \hat{\rho}_i^a}{\partial t}
 &+& \frac{\partial}{\partial r} ( \hat{\rho}_i^a v )
 + \frac{\partial}{\partial v} \left( \hat{\rho}_i^a c_i^a 
 \frac{v_0 - v}{T^a} \right) \nonumber \\
 &=& \frac{\tilde{\rho}_i^a(r,v,t)}{T_{\rm r}}
 [ \hat{P}_{0i}^a(v_0;r,t) - P_{0i}^a(v_0;r,t) ]
 + \frac{\partial^2}{\partial v^2} \left( \hat{\rho}_i^a
 \frac{A_i^a v^2}{T^a} \right) \nonumber \\
&=& \sum_b \sum_{i'} \int dv' \!\! \int\limits_{w < v'} \!\! dw \int dw_0 \,
 W_2^{ab}(v,i|v',i';w,i') \hat{\rho}_{i'}^a(r,v',v_0,t)
 \hat{\rho}_{i'}^b(r,w,w_0,t) \nonumber \\
 &-& \sum_b \sum_{i'} \int dv' \!\! \int\limits_{w < v} \!\! dw \int dw_0 \,
 W_2^{ab}(v',i'|v,i;w,i) \hat{\rho}_{i}^a(r,v,v_0,t)
 \hat{\rho}_{i}^b(r,w,w_0,t) \nonumber \\
 &+& \sum_{i' (\ne i)} W_1^a(i|i') \hat{\rho}_{i'}^a(r,v,v_0,t)
 - \!\sum_{i' (\ne i)} W_1^a(i'|i) \hat{\rho}_{i}^a(r,v,v_0,t) \nonumber \\
 &+& \hat{\nu}_i^{a+}(r,v,v_0,t) - \hat{\nu}_i^{a-}(r,v,v_0,t) \, .
\end{eqnarray}
Here, we have taken into account that the relaxation time $T^a$ and
the reasonable distribution of desired velocities $\hat{P}_{0i}^a$
may depend on driving style or vehicle type $a$. The specification of the
interaction rates $W_2^{ab}$ and the lane-changing rates $W_1^a$ is analogous
to the previous discussion (cf. (\ref{intrate}) and (\ref{w})).

\section{Derivation of macroscopic traffic equations}

The gas-kinetic traffic equations are not very suitable for computer
simulations since they contain too many variables. Moreover, the phase-space
densities are very small quantities and, therefore, subject to considerable
fluctuations so that a comparison with empirical data is difficult.
However, the special value of gas-kinetic traffic equations is that they
allow a systematic derivation of dynamic 
equations for the macroscopic (collective) quantities 
one is mainly interested in. 

\subsection{Definition of Variables}

The most relevant macroscopic quantities are the {\em vehicle densities}
\begin{equation}
 \rho_i(r,t) = \int dv \int dv_0 \, \hat{\rho}_i(r,v,v_0,t)
\end{equation}
and the {\em average velocities}
\begin{equation}
 V_i(r,t) \equiv \langle v \rangle_i = \int dv \, v P_i(v;r,t) 
\end{equation}
on lanes $i$. Here, we have applied the notation
\begin{equation}
 F_i(r,t) \equiv \langle f(v,v_0) \rangle_i = \int dv \int dv_0 \,
 f(v,v_0) \frac{\hat{\rho}_i(r,v,v_0,t)}{\rho_i(r,t)} 
\end{equation}
and introduced the {\em distribution of actual velocities}
\begin{equation}
 P_i(v;r,t) = \int dv_0 \, \frac{\hat{\rho}_i(r,v,v_0,t)}{\rho_i(r,t)}
 = \frac{\tilde{\rho}_i(r,v,t)}{\rho_i(r,t)}
\end{equation}
on lane $i$.
Analogous quantities can be defined for vehicles entering and leaving
the road at entrances and exits respectively.
\begin{equation}
 \nu_i^\pm(r,t) = \int dv \int dv_0 \, \nu_i^\pm(r,v,v_0,t)
\end{equation}
are the {\em rates of entering and leaving vehicles}, and
\begin{equation}
 V_i^\pm(r,t) \equiv \langle v \rangle_i^\pm = \int dv \, v
 P_i^\pm(v;r,t)
\end{equation}
their {\em average velocities}, where
\begin{equation}
 P_i^\pm(v;r,t) = \int dv_0 \, \frac{\hat{\nu}_i^{\pm}(r,v,v_0,t)}
 {\nu_i^\pm(r,t)}
\end{equation}
are the {\em velocity distributions} of entering and leaving 
vehicles respectively.
In addition, we will need 
the {\em velocity variance}
\begin{equation}
 \theta_i(r,t) \equiv \langle [v - V_i(r,t)]^2 \rangle_i
 = \int dv \, [v - V_i(r,t)]^2 P_i(v;r,t) = \langle v^2 \rangle_i
 - (\langle v \rangle_i)^2
\end{equation}
and the {\em average desired velocity}
\begin{equation}
 V_{0i}(r,t) = \int dv \int dv_0 \, v_0 \frac{\hat{\rho}_i(r,v,v_0,t)}
 {\rho_i(r,t)}
\end{equation}
on each lane $i$ as well as the {\em average interaction rate}
\begin{equation}
 \frac{1}{T_i^0} = \frac{1}{\rho_i(r,t)} 
 \int dv \, \tilde{\rho}_i(r,v,t) \!\!\int\limits_{w<v} \!\! dw \, ( v - w )
 \tilde{\rho}_i(r,w,t)
\end{equation}
of a vehicle on lane $i$ with other vehicles on the same lane.
\par
Finally, we will assume that the proportion $c_i$ of freely moving vehicles is
a function of the density $\rho_i$ on the respective lane $i$:
\begin{equation}
 c_i(r,v,v_0,t) \equiv c_i[\rho_i(r,t)] \, .
\end{equation}
The probabilities with which slower vehicles can immediately be overtaken
will mainly depend on the density on the neighboring 
lane $i\pm 1$, i.e.
\begin{equation}
 P_i^\pm \equiv P_i^\pm(\rho_{i\pm 1}) \, .
\end{equation}
A similar thing holds for the waiting times:
\begin{equation}
 \tilde{T}_i^\pm \equiv  \tilde{T}_i^\pm(\rho_{i\pm 1})
 \, , \qquad
 \hat{T}_i^\pm \equiv  \hat{T}_i^\pm(\rho_{i\pm 1}) \, .
\end{equation}
Consequently,
\begin{equation}
 p_i^\pm \equiv p_i^\pm(\rho_{i-1},\rho_i,\rho_{i+1};r,\not t)
\end{equation}
and
\begin{equation}
 T_i^\pm \equiv T_i^\pm(\rho_i,\rho_{i\pm 1};r,\!\not t) \, .
\end{equation}
From a theoretical standpoint, a generalization to more complicated 
functional dependences is easily possible, but their determination
from empirical data becomes very difficult, then.

\subsection{Derivation of Moment Equations}

We are now ready for deriving the desired macroscopic traffic equations
from the gas-kinetic equation 
(\ref{Boltz1}) with (\ref{acc}), (\ref{adj}), (\ref{VD}), (\ref{Boltz2}), (\ref{intrate}),
(\ref{spont}), and (\ref{w1}). Integration with
respect to $v_0$ gives us the {\em reduced gas-kinetic traffic equation}
\begin{mathletters}\label{red}\begin{eqnarray}
 \frac{\partial \tilde{\rho}_i}{\partial t}
 &+& \frac{\partial}{\partial r} (\tilde{\rho}_i v) + 
 \frac{\partial}{\partial v} \left( \tilde{\rho}_i c_i  
 \frac{\tilde{V}_{0i}(v) - v}{T} \right) 
 = \frac{\partial^2}{\partial v^2} \left( \tilde{\rho}_i
 \frac{A_i v^2}{T} \right) \label{reda} \\
 &-& (1-p_i) 
 \tilde{\rho}_i(r,v,t) \int dw \, (v - w) \tilde{\rho}_i(r,w,t)
 \label{redb} \\
 &+& p_{i-1}^+ \tilde{\rho}_{i-1}(r,v,t) \!\!\int\limits_{w < v} \!\! dw \, 
 (v - w) \tilde{\rho}_{i-1}(r,w,t) \label{redc} \\
 &+& p_{i+1}^- \tilde{\rho}_{i+1}(r,v,t) \!\!\int\limits_{w < v} \!\! dw \, 
 (v - w) \tilde{\rho}_{i+1}(r,w,t) \label{redd} \\
 &-& (p_i^+ + p_i^-) \tilde{\rho}_{i}(r,v,t) \!\!\int\limits_{w < v} \!\! dw
 \, (v - w) \tilde{\rho}_{i}(r,w,t) \label{rede} \\
 &+& \frac{1}{T_{i-1}^+} \tilde{\rho}_{i-1}(r,v,t)
 - \frac{1}{T_i^+} \tilde{\rho}_i(r,v,t) \label{redf} \\
 &+& \frac{1}{T_{i+1}^-} \tilde{\rho}_{i+1}(r,v,t)
 - \frac{1}{T_i^-} \tilde{\rho}_i(r,v,t) \label{redg} \\
 &+& \tilde{\nu}_i^+(r,v,t) - \tilde{\nu}_i^-(r,v,t) \label{redh}
\end{eqnarray}\end{mathletters}
with
\begin{equation}
 \tilde{V}_{0i}(v) \equiv \tilde{V}_{0i}(v;r,t)
 = \int dv_0 \, v_0 \frac{\hat{\rho}_i(r,v,v_0,t)}{\tilde{\rho}_i(r,v,t)}
\end{equation}
and
\begin{equation}
 \tilde{\nu}_i^\pm(r,v,t) = \int dv_0 \, \hat{\nu}_i^\pm(r,v,v_0,t) \, .
\end{equation}
In formula (\ref{red}), the deceleration term (\ref{redb}) 
stems from (\ref{intc}), 
the terms (\ref{redc}) to (\ref{rede}) reflecting immeditate overtaking come
from (\ref{inta}) and (\ref{intb}), and the lane-changing terms (\ref{redf}),
(\ref{redg}) originate from (\ref{w1}). 
The adaptation term $(\partial \hat{\rho}_i/\partial t)_{\rm ad}$
yields no contribution.
\par
We will now derive equations for the moments $\langle v^k \rangle$
by multiplying (\ref{red}) with $v^k$ and integrating with respect to $v$.
Due to
\begin{eqnarray}
 \int dv \, v^k \frac{\partial}{\partial v} \left( \tilde{\rho}_i c_i 
 \frac{\tilde{V}_{0i}(v) - v}{T} \right) &=& - \int dv \, k v^{k-1}
 \left( \tilde{\rho}_i c_i \frac{\tilde{V}_{0i}(v) - v}{T} \right)
\nonumber \\
 &=& - \frac{k c_i}{T} \rho_i ( \langle v^{k-1} v_0 \rangle_i - \langle v^k
 \rangle_i ) \, ,
\end{eqnarray}
\begin{equation}
 \int dv \, v^k \frac{\partial^2}{\partial v^2} \left( \tilde{\rho}_i
 \frac{A_i v^2}{T} \right) = k (k-1) \frac{\rho_i A_i}{T} \langle v^k \rangle_i
 \, ,
\end{equation}
and
\begin{equation}
 (1 - p_i) \int dv \, \tilde{\rho}_i(r,v,t) \int dw \, (wv^k - v^{k+1} )
 \tilde{\rho}_i(r,w,t) = (1-p_i) (\rho_i)^2 (\langle v \rangle_i
 \langle v^k \rangle_i - \langle v^{k+1} \rangle_i )
\end{equation}
we obtain the macroscopic moment equations
\begin{eqnarray}
 \frac{\partial}{\partial t} (\rho_i \langle v^k \rangle_i)
 + \frac{\partial}{\partial r} (\rho_i \langle v^{k+1} \rangle_i )
 &=& \frac{k c_i}{T} \rho_i \left[ \langle v^{k-1} v_0 \rangle_i - \langle v^k
 \rangle_i + (k-1)\frac{A_i}{c_i} \langle v^k \rangle_i \right] \nonumber \\
 &+& (1-p_i) (\rho_i)^2 (\langle v \rangle_i
 \langle v^k \rangle_i - \langle v^{k+1} \rangle_i ) \nonumber \\
 &+& \frac{p_{i-1}^+}{T_{i-1}^0} \rho_{i-1} [ \langle v^k
 \rangle_{i-1} + {\cal K}_{i-1}^k ] 
 - \frac{p_{i}^+}{T_{i}^0} \rho_{i} [ \langle v^k
 \rangle_{i} + {\cal K}_i^k ] \nonumber \\
 &+& \frac{p_{i+1}^-}{T_{i+1}^0} \rho_{i+1} [ \langle v^k
 \rangle_{i+1} + {\cal K}_{i+1}^k ]  
 - \frac{p_{i}^-}{T_{i}^0} \rho_{i} [ \langle v^k \rangle_{i} 
 + {\cal K}_i^k ] \nonumber \\
 &+& \frac{1}{T_{i-1}^+} \rho_{i-1} \langle v^k \rangle_{i-1} 
  - \frac{1}{T_{i}^+} \rho_{i} \langle v^k \rangle_{i} \nonumber \\
 &+& \frac{1}{T_{i+1}^-} \rho_{i+1} \langle v^k \rangle_{i+1} 
  - \frac{1}{T_{i}^-} \rho_{i} \langle v^k \rangle_{i} \nonumber \\
 &+& \nu_i^+(r,t) \langle v^k \rangle_i^+
  -  \nu_i^-(r,t) \langle v^k \rangle_i^- \, ,
\label{mom1}
\end{eqnarray}
where
\begin{equation}
 \langle v^k \rangle_i^\pm = \int dv \int dv_0 \, v^k 
 \frac{\hat{\nu}_i^\pm(r,v,v_0,t)} {\nu_i^\pm(r,t)}
= \int dv \, v^k \frac{\tilde{\nu}_i^\pm(r,v,t)}
 {\nu_i^\pm(r,t)}
\end{equation}
and
\begin{equation}
 {\cal K}_i^k = \frac{T_i^0}{\rho_i}
 \int dv \, \tilde{\rho}_i(r,v,t) \!\!\int\limits_{w<v} \!\! dw \, v^k (v-w) 
 \tilde{\rho}_i(r,w,t) - \langle v^k \rangle_i \, . 
\end{equation}
\par

\subsection{Fluid-Dynamic Multi-Lane Traffic Equations}

In order to derive dynamic equations for the densities
$\rho_i$ and average velocities $V_i$,
we need the relations
\begin{equation}
 \langle v^2 \rangle_i = \langle [ V_i + (v - V_i)]^2 \rangle_i
 = (V_i)^2 + 2 V_i \langle v - V_i \rangle_i + \langle (v-V_i)^2 \rangle_i
 = (V_i)^2 + \theta_i 
\end{equation}
and
\begin{equation}
 \rho_i \frac{\partial V_i}{\partial t} = \frac{\partial}{\partial t}
 ( \rho_i \langle v \rangle_i) - V_i \frac{\partial \rho_i}{\partial t} \, .
\end{equation}
Applying these and 
and using the abbreviations 
\begin{equation}
 \frac{1}{\tau_i^\pm} = \frac{p_i^\pm}{T_i^0} + \frac{1}{T_i^\pm} \, ,
\end{equation}
equation (\ref{mom1}) gives us the {\em density equations}
\begin{mathletters}\label{Dens}\begin{eqnarray}
 \frac{\partial \rho_i}{\partial t} + V_i \frac{\partial \rho_i}{\partial r}
 &=& - \rho_i \frac{\partial V_i}{\partial r} + \nu_i^+(r,t) - \nu_i^-(r,t)
 \label{Densa} \\
 &+& \frac{\rho_{i-1}}{\tau_{i-1}^+} - \frac{\rho_i}{\tau_i^+}
 + \frac{\rho_{i+1}}{\tau_{i+1}^-} - \frac{\rho_i}{\tau_i^-} \, .
\label{Densb}
\end{eqnarray}\end{mathletters}
This result is similar to previous multi-lane models.
In addition, we obtain the {\em velocity equations}
\begin{mathletters}\label{Veloc}\begin{eqnarray}
 \rho_i \frac{\partial V_i}{\partial t} + \rho_i V_i \frac{\partial V_i}
 {\partial r} &=& - \frac{\partial (\rho_i \theta_i) }{\partial r}
 + \frac{c_i \rho_i}{T} (V_{0i} - V_i) - (1-p_i) (\rho_i)^2 \theta_i 
\label{Veloca} \\
&+& \frac{\rho_{i-1}}{\tau_{i-1}^+} (V_{i-1} - V_i) 
 + \frac{\rho_{i+1}}{\tau_{i+1}^-} (V_{i+1} - V_i) \label{Velocb} \\
&+& \frac{p_{i-1}^+}{T_{i-1}^0} \rho_{i-1} {\cal K}_{i-1}^1
 + \frac{p_{i+1}^-}{T_{i+1}^0} \rho_{i+1} {\cal K}_{i+1}^1
 - \frac{p_i^+ + p_i^-}{T_{i}^0} \rho_{i} {\cal K}_{i}^1 \label{Velocd} \\
&+& \nu_i^+ (V_i^+ - V_i) - \nu_i^- (V_i^- - V_i ) \label{Velocc} 
 \end{eqnarray}\end{mathletters}
after some lengthy but straightforward calculations.
The terms containing the rates $\nu_i^+$ and $\nu_i^-$ reflect entering and
leaving vehicles, respectively.
Whereas the terms (\ref{Densa}) and (\ref{Veloca}) correspond
to the effects of vehicle motion, of acceleration towards the drivers' 
desired velocities,
and of deceleration due to interactions, the terms (\ref{Densb}),
(\ref{Velocb}), and (\ref{Velocd}) arise from overtaking and 
lane-changing maneuvers. (\ref{Velocb}) comes from differences between
the average velocities on neighboring lanes. In contrast, (\ref{Velocd})
originates from the fact that overtaking vehicles are, on average, somewhat
faster than the vehicles passed. Whereas (\ref{Velocd}) tends to produce 
differences between the average velocities of neighboring lanes, 
these are reduced by (\ref{Velocb}). The term (\ref{Velocc}) has a similar
form and interpretation like (\ref{Velocb}). It
is only negligible if entering vehicles are able 
to adapt to the velocities on the merging lane and exiting vehicles initially
have an average velocity similar to that on the lane which they 
are leaving so that $V_i^\pm \approx V_i$.
\par
In order to close equations (\ref{Dens}) and (\ref{Veloc}),
we must specify the interaction rates $1/T_i^0$, the correction terms
${\cal K}_i^1$, and the variances $\theta_i$. Utilizing that the 
empirical velocity distributions $P_i(v;r,t)$ are approximately
{\em normally distributed} \cite{Hel96,MuPi71,Pam55,Phi77}, we have
\begin{equation}
 P_i(v;r,t) \approx {\textstyle\frac{1}{\sqrt{2\pi \theta_i(r,t)}}}
 \mbox{e}^{-[v - V_i(r,t)]^2/[2\theta_i(r,t)]} 
\end{equation} 
which implies 
\begin{equation}
 \frac{1}{T_i^0} \approx \rho_i \sqrt{\frac{\theta_i}{\pi}}
\end{equation}
and
\begin{equation}
 {\cal K}_i^1 = \frac{\sqrt{\pi\theta_i}}{2} \, .
\end{equation}
With a detailled theoretical and empirical analysis it can be
shown \cite{Hel97} that the variances $\theta_i$ can be well
approximated by equilibrium relations of the form
\begin{equation}
 \theta_i =
 \frac{c_i(\rho_i){\cal C}_i(\rho_i,V_i) + A_i(\rho_i) (V_i)^2}
 {c_i(\rho_i) - A_i(\rho_i)} \, ,
\end{equation}
where ${\cal C}_i(\rho_i,V_i) = \langle (v - V_i)(v_0 - V_{0i}) \rangle_i^{\rm e}$ 
denotes the density- and velocity-dependent {\em equilibrium covariance}
between actual velocities $v$ and desired velocities $v_0$ on lane $i$. 
In the case of a speed limit there is $v_0 \approx V_{0i} \approx V_0$
so that ${\cal C}_i \approx 0$.
\par
For the average desired velocities $V_{0i}$ we have
\begin{equation}
 V_{0i} \equiv V_{0i}(r,t) \approx \hat{V}_{0i}(r,\!\not t)
\end{equation}
since
\begin{equation}
 \hat{P}_{0i}(v_0;r,\!\not t) - P_{0i}(v_0;r,t) 
 \approx 0
\label{adiab}
\end{equation}
due to the smallness of $T_{\rm r}$. 

\section{Effective macroscopic equations}

We will now investigate how the effective macroscopic traffic equations for
the total cross-section of the road look like. For this purpose we introduce
the {\em effective (average) density}
\begin{equation}
 \rho(r,t) = \frac{1}{I} \sum_i \rho_i(r,t) \, ,
\end{equation}
where $I \equiv I(r,\!\not t)$ is the {\em number of lanes} at place $r$,
and calculate the average $\overline{F}_i$ over all lanes for every
lane-specific quantity $F_i$ according to
\begin{equation}
 F \equiv \overline{F_i} = \sum_i \frac{\rho_i}{I \rho} F_i \, . 
\label{conven}
\end{equation}
This implies
\begin{equation}
  \rho F = \frac{1}{I} \sum_i \rho_i F_i \, .
\end{equation}
In addition we define the {\em effective rates} 
\begin{equation}
 \nu^\pm(r,t) = \frac{1}{I} \sum_i \nu_i^\pm(r,t) 
\end{equation}
{\em of entering and leaving vehicles}, respectively, and the corresponding
{\em lane-averages}
\begin{equation}
 F^\pm \equiv \overline{F_i^\pm} = \sum_i \frac{\nu_i^\pm}{I \nu^\pm} F_i^\pm \, .
\end{equation}
\par
In the following, we will apply the factorization approximation
\begin{equation}
 \overline{\rho_i F_i G_i} \approx \overline{\rho_i F_i} \, \overline{G_i}
 = \rho F G \, . 
\end{equation}
Summation of equation (\ref{mom1}) over $i$ and division by $I$
gives us
\begin{eqnarray}
 \frac{\partial}{\partial t} (\rho \overline{\langle v^k \rangle_i})
 + \frac{\partial}{\partial r} (\rho \overline{\langle v^{k+1} \rangle_i} )
 &=& \frac{k c}{T} \rho 
 [ \overline{\langle v^{k-1} v_0 \rangle_i} - \overline{\langle v^k
 \rangle_i} + (k-1) \overline{(A_i/c_i)} \, \overline{\langle v^k
 \rangle_i} ] \nonumber \\
 &=& (1-p) \rho^2 (\overline{\langle v \rangle_i
 \langle v^k \rangle_i} 
 - \overline{\langle v^{k+1} \rangle_i} ) \nonumber \\
 &+& \nu^+(r,t) \overline{\langle v^k \rangle_i^+}
  -  \nu^-(r,t) \overline{\langle v^k \rangle_i^-} \, ,
\label{mom3}
\end{eqnarray}
since the lane-changing terms cancel out each other: 
\begin{eqnarray}
\frac{1}{I} \sum_i \frac{1}{\tau_{i\mp 1}^\pm} \rho_{i\mp 1} 
\langle v^k \rangle_{i\mp 1} 
- \frac{1}{I} \sum_i \frac{1}{\tau_{i}^\pm} \rho_{i} \langle v^k \rangle_{i} 
= \frac{1}{I} \sum_{j} \frac{1}{\tau_{j}^\pm} \rho_{j} \langle v^k \rangle_{j} 
- \frac{1}{I} \sum_i \frac{1}{\tau_{i}^\pm} \rho_{i} \langle v^k \rangle_{i} 
= 0 \, .
\end{eqnarray}
A similar thing holds for the terms containing the corrections
${\cal K}_i^k$. As a consequence, only the terms describing 
vehicle motion, acceleration towards the drivers' desired velocities,
velocity fluctuations, deceleration due to vehicle interactions,
and entering or leaving vehicles are remaining. 
\par
Now, apart from the lane-average $\theta = \overline{\theta_i}$ 
of the velocity variances $\theta_i$
we additionally define the {\em total effective variance}
\begin{equation}
 \Theta = \overline{\langle (v - V)^2 \rangle_i}
 = \overline{\langle [(v - V_i) + (V_i - V)]^2\rangle_i}
 = \theta + \overline{(V_i - V)^2} \, .
\end{equation}
Applying this and relation 
\begin{eqnarray}
 \overline{\langle v^2 \rangle_i} &=& \overline{(V_i)^2 + \theta_i}
 = \overline{[V + (V_i - V)]^2} + \theta \nonumber \\
 &=& V^2 + \overline{(V_i - V)^2} + \theta = V^2 + \Theta
\end{eqnarray}
to equation (\ref{mom3}),
after some lengthy but
straightforward calculations we obtain the {\em effective density equation}
\begin{equation}
 \frac{\partial \rho}{\partial t} + V \frac{\partial \rho}{\partial r}
 = - \rho \frac{\partial V}{\partial r} + \nu^+ - \nu^- 
\label{effdens}
\end{equation}
and the {\em effective velocity equation}
\begin{eqnarray}
 \rho \frac{\partial V}{\partial t} + \rho V \frac{\partial V}{\partial r}
 &=& - \frac{\partial (\rho\Theta)}{\partial r} + \frac{c \rho}{T}
 (V_0 - V) - (1-p) \rho^2 \theta \nonumber \\
 &+& \nu^+ (V^+ - V) - \nu^- (V^- - V) \, .
\label{effveloc}
\end{eqnarray}

\subsection{Comparison with former macroscopic traffic models}

Taking a look at the effective density equation (\ref{effdens}), we recognize that
it has the form of a {\em continuity equation}. This reflects the conservation
of the number of vehicles, i.e. no vehicle on the considered road is produced
or gets lost. The continuity equation was first proposed in the fluid-dynamic
traffic model by Lighthill and Whitham \cite{LiWhi55,Whi74}, and it is also a 
component of most other macroscopic traffic models 
\cite{Pay71,Kue84,KueRoe91,KeKo93,KeKo94,Hel95a,Hel95b,Hel96}.
\par
The effective velocity equation (\ref{effveloc}) can alternatively 
be represented in the form
\begin{equation}
 \rho \frac{\partial V}{\partial t} + \rho V \frac{\partial V}{\partial r}
 = - \frac{\partial {\cal P}}{\partial r} + \frac{\rho}{\tau}
 (V_{\rm e} - V) + \nu^+ (V^+ - V)
 - \nu^- (V^- - V)  \, , 
\label{all}
\end{equation}
where we have introduced the so-called
{\em traffic pressure} \cite{PrHe71,Phi77,Phi79}
\begin{equation}
 {\cal P} = \rho \Theta \, .
\end{equation}
The {\em effective relaxation time} 
\begin{equation}
 \tau(\rho)  = \frac{T}{c(\rho)}
\end{equation}
increases with decreasing effective
proportion $c(\rho)$ of freely moving vehicles and reflects the 
well-known ``frustration
effect'' that the average relaxation time increases with growing
vehicle density $\rho$. In addition, we have
introduced the {\em equilibrium velocity}
\begin{equation}
 V_{\rm e} = V_0 - \tau (1-p) \rho \theta
\label{eqvel}
\end{equation}
which is determined by the average desired velocity $V_0$, 
diminished by a term 
arising from deceleration maneuvers due to vehicle interactions.
The velocity equations proposed by Prigogine et al. \cite{PrHe71},
Paveri-Fontana \cite{Pa75}, Phillips \cite{Phi77,Phi79}, Payne \cite{Pay71}, 
as well as Kerner and Konh\"auser \cite{KeKo93} 
can all be written in the form (\ref{all}) \cite{Hel96}. 
However, the specification of the functions ${\cal P}$, $\tau$, and
$V_{\rm e}$ is varying from one model to another. The terms
$\nu^\pm (V^\pm - V)$ have been
neglected by all previous models, despite the fact that entrances (and exits)
reduce the freeway capacity due to $V^\pm \le V$. Moreover, the effective
gas-kinetic models proposed by Prigogine et al. \cite{PrHe71},
Paveri-Fontana \cite{Pa75}, and Phillips \cite{Phi77,Phi79} imply 
$V_{\rm e} = V_0 - \tau(1-p) \rho \Theta$ instead of relation
(\ref{eqvel}). This is a consequence of the fact that they treat
the total freeway section like a {\em single} lane of greater capacity and
possibilities for overtaking. However, the approximation $\Theta \approx
\theta$ is only valid for $V_i \approx V$. Although it is normally justified
for American freeways, on European autobahns it is only valid at densities
greater than 35 vehicles per kilometer and lane \cite{Hel97}.

\section{Summary and Outlook}

In this paper we have derived a macroscopic traffic model for uni-directional
multi-lane roads. Our considerations started from plausible assumptions about
the behavior of driver-vehicle units regarding acceleration, 
deceleration, velocity fluctuations, overtaking,
and lane-changing maneuvers. In addition, we distinguished
freely moving and queuing vehicles since these behave differently. The 
resulting gas-kinetic traffic model is a generalization of Paveri-Fontana's
Boltzmann-like traffic equation. It can be extended to situations where
different vehicle types or driving styles are to be investigated.
\par
The gas-kinetic traffic equations not only allow to derive dynamic equations for
the vehicle density on each lane, but also for the
average velocity. In addition, we obtained {\em effective} macroscopic
traffic equations for the total cross-section of the road from the
multi-lane model. The resulting equations implied corrections with respect to
previous traffic models, but agree with the ones by
Prigogine et al. and Paveri-Fontana in the special case $V_i \approx
V \approx V^\pm$. 
\par
Strictly speaking, the presented model is only valid for small vehicle
densities. However,  with the methods discussed in Refs.~\cite{Hel96,Hel97}
it can be easily generalized to a traffic model for arbitrary densities.
Since the calculations are rather lengthy but straightforward, they
have been omitted, here.
\par
Present research focuses on the implementation of the multi-lane model on a computer
with the objective of carrying out numerical traffic simulations. The functional
relations regarding the probabilities of overtaking, the proportions of freely
moving vehicles, the waiting times for lane-changing maneuvers, etc. are
evaluated on the basis of empirical data. Some of the questions we
are going to investigate by means of multi-lane simulations are the
following:
\begin{enumerate}
\item In which way does on-ramp traffic influence and destabilize the traffic
flow on the other lanes? How does the destabilization effect depend on the
traffic volume, the length of the on-ramp lane, the total lane number, etc.?
\item In case of a reduction of the number of lanes, is it better to close the
left-most or the right-most lane? 
\item Is the organization of American freeways or of European autobahns more
efficient, or is it a suitable mixture of both? Remember that American
freeways are characterized by uniform speed limits and the fact that
overtaking as well as lane changing is allowed on both neighboring lanes.
In contrast, on European autobahns 
often no speed limit is prescribed (at least in
Germany) and average velocity normally increases with growing lane number
since overtaking is only allowed on the left-hand lane.
\item In which traffic situations do stay-in-lane recommendations increase the
efficiency of roads?
\end{enumerate}


\begin{thebibliography}{10}

\bibitem{LiWhi55}
M.~J. Lighthill and G.~B. Whitham (1955) 
{\it Proceedings of the Royal Society A} {\bf 229}, 317.

\bibitem{Ri56}
P.~I. Richards (1956) {\it Operations Research} {\bf 4}, 42.

\bibitem{Pay71}
H.~J. Payne (1971) In: G.~A. Bekey, ed. 
{\it Mathematical Models of Public Systems}, Vol. 1.
(Simulation Council, La Jolla, CA).

\bibitem{Pay79b}
H.~J. Payne (1979) 
{\it Transportation Research Record} {\bf 722}, 68.

\bibitem{Kue84}
R.~D. K{\"u}hne (1984) 
In: I. Volmuller and R. Hamerslag, eds. {\it Proceedings of the 9th
  International Symposium on Transportation and Traffic Theory}.
(VNU Science Press, Utrecht, The Netherlands).

\bibitem{KueRoe91}
R.~D. K{\"u}hne and M.~B. R{\"o}diger (1991) 
In: B.~L. Nelson, W.~D. Kelton,  and G.~M. Clark, eds. {\it Proceedings of the
  1991 Winter Simulation Conference}.
(Society for Computer Simulation International, Phoenix, Arizona).

\bibitem{KeKo93}
B.~S. Kerner and P. Konh{\"a}user (1993) 
{\it Physical Review E} {\bf 48}, 2335.

\bibitem{KeKo94}
B.~S. Kerner and P. Konh{\"a}user (1994) 
{\it Physical Review E} {\bf 50}, 54.

\bibitem{Hel95a}
D. Helbing (1995) {\it Physica A} {\bf 219}, 375.

\bibitem{Hel95b}
D. Helbing (1995) {\it Physica A} {\bf 219}, 391.

\bibitem{Hel96}
D. Helbing (1996) Derivation and empirical validation of a refined traffic
flow model. {\it Physica A} {\bf 233}, 253.

\bibitem{KueBe93}
R.~D. K{\"u}hne and R. Beckschulte (1993) 
In: C.~F. Daganzo, ed. {\it Proceedings of the 12th International Symposium on the
  Theory of Traffic Flow and Transportation}.
(Elsevier, Amsterdam, The Netherlands).

\bibitem{Hel95c}
D. Helbing (1995) {\it Physical Review E} {\bf 51}, 3164.

\bibitem{Kue87}
R.~D. K{\"u}hne (1987) 
In: N.~H. Gartner and N.~H.~M. Wilson, eds. {\it Proceedings of the 10th
  International Symposium on Transportation and Traffic Theory}.
(Elsevier, New York).

\bibitem{KueKr92}
R.~D. K{\"u}hne and A. Kroen (1992) Knowledge-based optimization of line
control systems for freeways.
(Steierwald Sch{\"o}nharting \& Partner, Beratende Ingenieure, Stuttgart,
Germany).

\bibitem{Hel95d}
D. Helbing (1995) 
In: M. Snorek, M. Sujansky, and A. Verbraeck, eds.
{\it Modelling and Simulation 1995}.
(The Society for Computer Simulation International, Instanbul, Turkey).

\bibitem{HiReWe93}
M. Hilliges, R. Reiner,  and W. Weidlich (1993) 
In: A. Pave, ed. {\it Modelling and Simulation 1993}.
(Society for Computer Simulation International, Ghent, Belgium).

\bibitem{Hil95}
M. Hilliges (1995) A phenomenological model for dynamic traffic flow in
  networks.
{\it Transportation Research B} {\bf 29}, 407.

\bibitem{Hil96}
M. Hilliges and N. Koch (1996) In: D. E. Wolf, M. Schreckenberg, and
A. Bachem, eds. {\it Traffic and Granular Flow}.
(World Scientific, Singapore).

\bibitem{Hel97}
D. Helbing {\it Verkehrsdynamik. Neue physikalische
  Modellierungskonzepte.} (Springer, Berlin, 1997).

\bibitem{Bra68}
D. Braess (1968) {\it Unternehmensforschung} {\bf 12}, 258.

\bibitem{Bas92}
T. Bass (May/1992) Road to ruin.
{\it Discover,} 56.

\bibitem{Kno93}
A. Knop (March/1993) {\it Natur}, 76.

\bibitem{GaHeWe62}
D.~C. Gazis, R. Herman,  and G.~H. Weiss (1962) 
{\it Operations Research} {\bf 10}, 658.

\bibitem{MuPi71}
P.~K. Munjal and L.~A. Pipes (1971) 
{\it Transportation Research} {\bf 5}, 241.

\bibitem{MuHs71}
P.~K. Munjal, Y.-S. Hsu,  and R.~L. Lawrence (1971) 
{\it Transportation Research} {\bf 5}, 257.

\bibitem{Ror76}
J. R{\o}rbech (1976) 
{\it Transportation Research Record} {\bf 596}, 22.

\bibitem{MaNa81}
Y. Makigami, T. Nakanishi, M. Toyama,  and R. Mizote (1983) 
In: V. F. Hurdle, E. Hauer and G. N. Stewart, eds. 
{\it Proceedings of the 8th International Symposium on Transportation and
  Traffic Theory}. (University of Toronto Press, Toronto, Ontario).

\bibitem{MiBe84}
P.~G. Michalopoulos, D.~E. Beskos,  and Y. Yamauchi (1984) 
{\it Transportation Research B} {\bf 18}, 377.

\bibitem{HaHu79}
E. Hauer and V.~F. Hurdle (1979) 
{\it Transportation Research Record} {\bf 722}, 75.

\bibitem{Pay79}
H.~J. Payne (1979) In: W.~S. Levine, E. Lieberman,  and J.~J. Fearnsides, eds. {\it Research
  Directions in Computer Control of Urban Traffic Systems}.
(American Society of Civil Engineers, New York).

\bibitem{Pap83}
M. Papageorgiou (1983) {\it Applications of Automatic Control Concepts to
  Traffic Flow Modeling and Control}.
(Springer, Heidelberg, Germany).

\bibitem{Cre85}
M. Cremer and A.~D. May (1985) {\it An Extended Traffic Model for Freeway
  Control}.
(Research Report UCB-ITS-RR-85-7.
Institute of Transportation Studies, University of California, Berkeley).

\bibitem{Smu87}
S.~A. Smulders (1987) 
In: N.~H. Gartner and N.~H.~M. Wilson, eds. {\it Proceedings of the 10th
  International Symposium on Transportation and Traffic Theory}.
(Elsevier, New York).

\bibitem{RaLi87}
A.~K. Rathi, E.~B. Lieberman,  and M. Yedlin (1987) 
{\it Transportation Research Record} {\bf 1112}, 61.

\bibitem{Dag75}
C.~F. Daganzo (1975) {\it Transportation Research} {\bf 9}, 339.

\bibitem{And1}
F.~C. Andrews (1970) {\it Transportation Research} {\bf 4}, 359.

\bibitem{And2}
F.~C. Andrews (1970) {\it Transportation Research} {\bf 4}, 367.

\bibitem{And3}
F.~C. Andrews (1973) {\it Transportation Research} {\bf 7}, 223.

\bibitem{And4}
F.~C. Andrews (1973) {\it Transportation Research} {\bf 7}, 233.

\bibitem{Be79}
A.~E. Beylich (1979) In: R. Campargue, ed. {\it Rarefied Gas Dynamics}, 
Vol. 1. (Commissariat a l'Energie Atomique, Paris, France).

\bibitem{Be81}
A.~E. Beylich (1981) In: G. Adomeit and H.-J. Frieske, eds. 
{\it {N}eue {W}ege in der {M}echanik}.
(VDI-Verlag, D{\"u}sseldorf, Germany).

\bibitem{Lam78}
M. Lampis (1978) {\it Transportation Science} {\bf 12}, 16.

\bibitem{PrAn60}
I. Prigogine and F.~C. Andrews (1960) {\it Operations Research} {\bf 8}, 789.

\bibitem{Pri61}
I. Prigogine (1961) In: R. Herman, ed. {\it Theory of Traffic Flow}.
(Elsevier, Amsterdam, The Netherlands).

\bibitem{PrHe71}
I. Prigogine and R. Herman (1971) {\it Kinetic Theory of Vehicular Traffic}.
(Elsevier, New York).

\bibitem{Pa75}
S.~L. Paveri-Fontana (1975) {\it Transportation Research} {\bf 9}, 225.

\bibitem{AlBe78}
E. Alberti and G. Belli (1978) {\it Transportation Research} {\bf 12}, 33.

\bibitem{Pam55}
F. Pampel (1955) {\it Ein {B}eitrag zur {B}erechnung der
  {L}eistungsf{\"a}higkeit von {S}tra{\ss}en}.
(Kirschbaum, Bielefeld, Germany).

\bibitem{QSoz}
D. Helbing (1995) {\it Quantitative Sociodynamics. {S}tochastic Methods and
  Models of Social Interaction Processes}.
(Kluwer Academic, Dordrecht, The Netherlands).

\bibitem{Whi74}
G.~B. Whitham (1974) {\it Linear and Nonlinear Waves}.
(Wiley, New York).

\bibitem{Phi77}
W.~F. Phillips (1977) {\it Kinetic {M}odel for {T}raffic {F}low}.
(Report No. DOT/RSPD/ DPB/50--77/17.
National Technical Information Service, Springfield, VA 22161).

\bibitem{Phi79}
W.~F. Phillips (1979) {\it Transportation Planning and Technology} 
{\bf 5}, 131.

\end{thebibliography}
\end{document}